\documentclass[12pt]{iopart}
\usepackage{amssymb,amsbsy}
\usepackage{iopams}
\usepackage{graphicx}
\usepackage{dcolumn}
\usepackage[usenames]{color}

\usepackage{bm}
\usepackage{amsgen,amsfonts}

\newcommand{\Lapse}{\alpha}       
\newcommand{\Shift}{\beta}   

\newcommand{\TrExCurv}{K}    

\newcommand{\dtime}{\partial_t}   

\newcommand{\CF}{\psi}              

\newcommand{\CLapse}{\tilde{\alpha}}     
\newcommand{\dtCMetric}{\tilde{u}}  

\newcommand{\CA}{\tilde{A}}         
\newcommand{\CRicciS}{\tilde{R}}    

\newcommand{\CCD}{{\tilde\nabla}\!} 
\newcommand{\CCDu}{{\tilde\nabla}}  
\newcommand{\CLong}[1]{(\tilde{\mathbb L}{#1})} 

\newcommand{\CTwoMetric}{\tilde{h}}         
\newcommand{\CSpatialNormal}{\tilde{s}}

\newcommand{\OmegaOrbitID}{\Omega_0}  

\begin{document}

\title[Reducing spurious gravitational radiation by using 
conformally curved initial data]
{Reducing spurious gravitational radiation 
in binary-black-hole 
simulations by using conformally curved initial data}
\author{Geoffrey Lovelace}
\address{Center for Radiophysics and Space Research, Cornell University, Ithaca, NY, 14853}
\address{Theoretical Astrophysics 130-33, California Institute of Technology, Pasadena, CA 91125}\ead{geoffrey@astro.cornell.edu}

\begin{abstract}
At early times in numerical evolutions of binary black holes, current simulations contain an initial burst of spurious gravitational radiation (also called ``junk radiation'') which is not astrophysically realistic. The spurious radiation is a consequence of how the binary-black-hole initial data are constructed: the initial data are typically assumed to be conformally flat. In this paper, I adopt a curved conformal metric that is a superposition of two boosted, non-spinning black holes that are approximately 15 orbits from merger. I compare junk radiation of the superposed-boosted-Schwarzschild (SBS) initial data with the junk of corresponding conformally flat, maximally sliced (CFMS) initial data. The SBS junk is smaller in amplitude than the CFMS junk, with the junk's leading-order spectral modes typically being reduced by a factor of order two or more.
\end{abstract}
\pacs{04.25.D-,04.25.dg,04.20.Ex,02.70.Hm}

\section{Introduction}

One of the most important sources of gravitational waves for LIGO~\cite{Barish:1999} is the inspiral and merger of two black holes. LIGO has reached its design sensitivity and can detect stellar-mass binary-black-hole mergers as distant as about 100 megaparsecs \cite{Cutler2002}. 

 
The gravitational waveform of a binary-black-hole merger cannot be computed using analytic techniques but must be obtained by solving the Einstein equations numerically. Currently, simulations 
of binary black holes are based on splitting the four-dimensional spacetime into a series of three-dimensional spatial slices; to start an evolution, one must construct initial data for the first slice. This initial data must i) represent the desired physical situation (i.e., two black holes about to merge), and ii) satisfy the vacuum Einstein constraint equations:
\begin{eqnarray}\label{junkeq:constraints}
G_{nn}=0,\nonumber\\
G_{nj}=0,
\end{eqnarray} where the subscript $n$ refers to components normal to the initial slice and the subscript $j$ refers to components tangent to the slice. The normal-normal and normal-spatial equations are called the Hamiltonian and momentum constraints, respectively. The vacuum Einstein evolution equations, \(G_{ij}=0\), are solved to step from the initial slice to subsequent slices.

There are several methods that generate constraint-satisfying initial data (for a review, see, e.g.,~\cite{Cook2000}), including 
Bowen-York puncture data~\cite{bowenyork80,Brandt1997} and quasiequilibrium excision data~\cite{Grandclement2001,Gourgoulhon2001,Cook2002,Cook2004,Caudill-etal:2006}. However, these methods generally assume that the initial spatial metric \(g_{ij}\) is conformally flat:
\begin{eqnarray}
g_{ij} = \psi^4 f_{ij},
\end{eqnarray} where $f_{ij}$ is the metric of flat space. This simplifying assumption causes spurious gravitational radiation (also called ``junk radiation'') to be present in the early phases of the simulation. Specifically, it is known that a stationary, isolated black hole with linear \cite{YorkJunk} or angular \cite{GaratPriceJunk} momentum cannot be sliced so that the spatial metric is conformally flat. Attempting to construct constraint-satisfying, conformally flat initial data for boosted or spinning black holes yields holes that are not in equilibrium but are unphysically perturbed. As they relax to an equilibrium configuration, gravitational waves are emitted.

In general, the black holes in a binary have both linear and angular momentum; therefore, binary-black-hole simulations using conformally flat initial data will also contain junk radiation. Before one can extract the physically relevant gravitational wave signal, one must first evolve the unphysical system until the spurious waves have left the computational domain; thus 
junk 
radiation adds to the already considerable computational 
expense of the simulation. But besides the additional cost, junk radiation causes several other undesirable effects. Spurious gravitational radiation can
unrealistically shorten the time until the black holes 
merge~\cite{bodeEtAl:2008}. Junk radiation also makes accurate comparison with post-Newtonian waveforms more difficult: it reduces the accuracy of the simulations and postpones the starting time at which the post-Newtonian comparison can begin~\cite{boyleEtAl2007,HannamEtAl:2008}. 
For simulations of binary black holes which recoil, the 
spurious gravitational radiation and astrophysically-realistic 
gravitational waves both carry linear momentum in the direction of the kick; consequently, before integrating the 
radiated momentum flux to obtain the
kick velocity, one must first wait for the junk radiation to leave the computational 
grid~\cite{Choi-Kelly-Boggs-etal:2007,gonzalezEtAl:2007}. 

To minimize these effects,
it is desirable to reduce the spurious radiation as much as possible. 
Because conformal flatness is known to contribute to the 
junk radiation, one common approach is to choose a \emph{curved} conformal metric; suitable choices lead to smaller amounts of junk radiation. (Note, however, that even conformally flat initial data can be constructed such that less 
spurious radiation is emitted than for standard Bowen-York puncture 
data~\cite{DainEtAl:2002}.) In Ref.~\cite{ReducingPunctureJunk}, the authors 
(building on the work in Refs.~\cite{Dain2001a,Dain2001b,KrivanPrice}) have constructed 
and evolved conformally 
curved initial data to reduce the amount of junk radiation in head-on mergers of spinning holes; by using a conformal metric that is a superposition of two spinning black holes, they find that the leading order, quadrupole spherical-harmonic mode of the spurious radiation is reduced. Furthermore, Ref.~\cite{Yunes2006a} proposes using conformal metrics made by asymptotically matching post-Newtonian and perturbed-black-hole metrics, and initial data based on post-Newtonian free data have been 
constructed~\cite{Tichy2002,Blanceht2003,Nissanke2006,Kellyetal2007}; however, to the best of my knowledge, no numerical simulations evolving these initial data have been published to date. 

In this paper, I construct conformally curved initial data for two nonspinning black holes of equal mass in orbits with low eccentricity.  In particular, I use a conformal metric [Eq.~(\ref{junkeq:SBSMetric})] that is a superposition of two boosted Schwarzschild black holes. To construct binary-black-hole initial data with nearly-extremal spins, Ref.~\cite{Lovelace2008} adopts a conformal metric similar to the one used in this paper but superposes two boosted, spinning, Kerr-Schild black holes instead of two nonspinning, boosted, Schwarzschild black holes. Note that the superposed metrics in this paper and in Ref.~\cite{Lovelace2008} are similar to the superposed-Kerr-Schild conformal metrics used in Refs.~\cite{Matzner1999,Marronetti-Matzner:2000,MM2000}; however, the conformal metrics used here and in Ref.~\cite{Lovelace2008} are flat everywhere except near each black hole.



After choosing superposed-boosted-Schwarzschild free data, I then combine them with quasiequilibrium boundary conditions developed by Cook~\cite{Cook2002}, Cook and Pfeiffer~\cite{Cook2004}, and Caudill~et.~al.~\cite{Caudill-etal:2006}. I solve the constraint equations using the Caltech-Cornell pseudospectral elliptic solver~\cite{Pfeiffer2003}. After reducing the eccentricity of the holes' orbits by using the technique of Pfeiffer~et.~al.~\cite{EccentricityPaper} (which was extended to the conformally curved case in Ref.~\cite{Lovelace2008}), I evolve the holes using the Caltech-Cornell code~\cite{Lindblom2006,Scheel2006}.

The remainder of this paper is organized as follows. In Sec.~\ref{junksec:XCTSDeriv}, I summarize the formalism that I use to solve the initial value problem. In Sec.~\ref{junksec:SBSMetric}, I describe how to construct initial data whose conformal metric is a superposition of two boosted Schwarzschild black holes. In Sec.~\ref{junksec:datasets}, I choose a conformally flat, maximally sliced (CFMS) initial data set and a superposed-boosted-Schwarzschild (SBS) initial data set that is physically comparable. In particular, I use the eccentricity-reduction technique of~\cite{EccentricityPaper} (originally developed under the assumption of conformal flatness) so that sets CFMS and SBS both have very little orbital eccentricity. Evolutions of the 
conformally curved data are discussed in Sec.~\ref{sec:Evolutions}, and the junk radiation of sets CFMS and SBS are compared in Sec.~\ref{junksec:junkcompare}. A brief conclusion is made in Sec.~\ref{junksec:conclusion}.

\section{The initial value problem}
\label{junksec:XCTSDeriv}
\subsection{The constraint equations}
To construct constraint-satisfying initial data, I begin with the usual 3+1 split, in which the four-dimensional spacetime, with metric \(g_{\mu\nu}\), is split into a series of three-dimensional spatial slices with spatial metric \(g_{ij}\). The spacetime metric \(g_{\mu\nu}\) is related to the spatial metric \(g_{ij}\), the lapse \(\alpha\), and the shift \(\beta^i\) by
\begin{equation}
ds^2 = g_{\mu\nu}dx^\mu dx^\nu = -\alpha^2 dt^2 + g_{ij}(dx^i+\beta^i dt)(dx^j+\beta^j dt).
\end{equation} Here and throughout the rest of this paper, the Einstein summation convention is assumed. Greek indices refer to spacetime coordinates and are raised and lowered with the spacetime metric \(g_{\mu\nu}\) and its inverse. Latin indices refer to spatial coordinates of a \(t=const\) slice and are raised and lowered with the spatial metric \(g_{ij}\) and its inverse.

On the initial ($t=0$) slice, the initial data must specify \(g_{ij}\) and the extrinsic curvature \(K_{ij}\), which is essentially the rate of change of \(g_{ij}\) in the normal direction. The extrinsic curvature is related to the time derivative of the metric \(\partial_t g_{ij}\) and to the lapse and shift by
\begin{equation}
\partial_t g_{ij} = -2 \alpha K_{ij} + 2\nabla_{(i}\beta_{j)}.
\end{equation} The initial values of \(g_{ij}\) and \(K_{ij}\) must be chosen so that i) the solution contains the desired physical content, and ii) the constraint equations~(\ref{junkeq:constraints}) are satisfied.

A systematic way to solve these equations is given by the extended conformal thin sandwich (XCTS) formalism \cite{York1999,PfeifferYork2003}. In this formalism, one expands \(g_{ij}\) and \(K_{ij}\) as follows:
\begin{eqnarray}
g_{ij} = \psi^4 \tilde{g}_{ij},\nonumber\\
K_{ij} = A_{ij} + \frac{1}{3}g_{ij} K.
\end{eqnarray} Then, one chooses the conformal metric \(\tilde{g}_{ij}\), the trace of the extrinsic curvature \(K\), and the time derivatives of both, \(\tilde{u}_{ij}:=\partial_t \tilde{g}_{ij}\) and \(\partial_t K\). The constraint equations (\ref{junkeq:constraints}) are then reduced to elliptic equations for the conformal factor \(\psi\) and the shift \(\beta^i\). A fifth elliptic equation for \(\alpha\psi\) determines the lapse; it is not a constraint, but appears because the free data include \(\partial_t K\) instead of \(\tilde{\alpha}:=\psi^{-6}\alpha\). (Alternatively, one could use the ``standard'' conformal thin sandwich equations~\cite{York1999,PfeifferYork2003}, in which the free data are \(\tilde{g}_{ij}\), \(\tilde{u}_{ij}\), \(K\), and \(\tilde{\alpha}\).)

Together, these equations form a second-order, nonlinear, coupled elliptic system called the extended conformal thin sandwich (XCTS) equations, which are (e.g., Eq.~(8) of~\cite{EccentricityPaper}):
\numparts
\begin{eqnarray}\label{junkeq:XCTSa}
\CCDu^2\CF-\frac{1}{8}\CRicciS\CF-\frac{1}{12}\TrExCurv^2\CF^5 
+\frac{1}{8}\CF^{-7}\CA^{ij}\CA_{ij} = 0,\\
\label{junkeq:XCTSb}
\CCD_j\Big(\frac{\psi^7}{2(\alpha\psi)}\CLong{\Shift}^{ij}\Big)
-\frac{2}{3}\CF^6\CCDu^i\TrExCurv
-\CCD_j\Big(\frac{\psi^7}{2(\alpha\psi)}\dtCMetric^{ij}\Big) = 0,\\
\CCDu^2(\Lapse\CF)-(\Lapse\CF)\bigg[\frac{\CRicciS}{8}\!+\!\frac{5}
{12}\TrExCurv^4\CF^4\!
+\!\frac{7}{8}\CF^{-8}\CA^{ij}\CA_{ij}\bigg]
\label{junkeq:Lapse2}
=-\CF^5(\dtime\TrExCurv-\Shift^k\partial_k\TrExCurv).
\label{junkeq:XCTSc}
\end{eqnarray}
\endnumparts Here \(\tilde{\nabla}\) is the gradient with respect to \(\tilde{g}_{ij}\), the ``longitudinal operator'' \(\tilde{\mathbb{L}}\) is twice the symmetric, trace-free gradient (i.e., the ``shear'') with respect to \(\tilde{g}_{ij}\), i.e.,
\begin{equation}
(\tilde{\mathbb{L}}V)_{ij} := \tilde{\nabla}_i V_j + \tilde{\nabla}_j V_i - \frac{2}{3} \tilde{g}_{ij} \tilde{\nabla}_k V^k,
\end{equation} and
\begin{equation}
\tilde{A}^{ij} = \psi^{10} A^{ij} = \frac{\psi^7}{2 (\alpha\psi)}\left[(\tilde{\mathbb{L}}\beta)^{ij}-\tilde{u}^{ij}\right].
\end{equation}

The initial value problem now amounts to i) choosing the free data (
\(\tilde{g}_{ij}\), \(\tilde{u}_{ij}\), \(K\), and \(\partial_t K\)), ii) choosing boundary conditions for \(\psi\), \(\alpha\psi\), and 
\(\beta^i\), and iii) solving Eqs.~(\ref{junkeq:XCTSa})--(\ref{junkeq:XCTSc}) for \(\psi\), \(\alpha\psi\), and \(\beta^i\). Most of these have preferred choices, motivated by the requirement that, in the comoving coordinates, the initial data contain two black holes at rest. These \emph{quasiequilibrium} conditions will be discussed in the next subsection. The remaining quantities will be dealt with in Sec.~\ref{junksec:SBSMetric}.

\subsection{Quasiequilibrium free data and boundary conditions}
In the XCTS formalism described in the previous subsection, the physical content of the data is selected by the choice of both the free data (\(\tilde{g}_{ij}\), \(K\), and their time derivatives) and by the boundary conditions. We wish to make choices that represent the physical situation of two (otherwise isolated) black holes orbiting each other. In the quasiequilibrium method \cite{Grandclement2001, Gourgoulhon2001,Cook2002,Cook2004,Caudill-etal:2006,EccentricityPaper} used in this paper, there are preferred choices for many of the free data and boundary conditions. 

\subsubsection{Free data}
In quasiequilibrium initial data, the coordinates are required to (initially) be \emph{comoving} with the black holes. If the holes are also in equilibrium, time derivatives in the comoving frame should initially be small. Quasiequilibrium initial data therefore choose
\begin{eqnarray}
\tilde{u}_{ij} = 0,\\
\partial_t K = 0.
\end{eqnarray} The remaining free data, \(\tilde{g}_{ij}\) and \(K\), can be chosen freely. In Sec.~\ref{junksec:SBSMetric}, I make particular choices for \(\tilde{g}_{ij}\) and \(K\).

\subsubsection{Outer boundary conditions}
The computational domain can be represented by only a finite number of gridpoints, so it necessarily will have an outer boundary \(\mathcal{B}\), which here is taken to be a coordinate sphere whose radius \(R\) is so much larger than all other length scales that it is effectively ``infinitely far away.'' (In practice, the outer boundary is roughly \(10^9\) times larger than the size of each black hole.) The physical requirement that the binary is isolated (i.e., that the spacetime is asymptotically flat) corresponds to the conditions
\begin{eqnarray}
\psi = 1 \mbox{ on }\mathcal{B},\label{junkeq:outerBCpsi}\\ 
\alpha = 1 \mbox{ on }\mathcal{B}\label{junkeq:outerBClapsePsi}, 
\end{eqnarray} provided that \(\tilde{g}_{ij}\) is asymptotically flat. 

The outer boundary condition on the shift is set by the requirement that the coordinates are initially comoving with the black holes. Therefore, in the asymptotically flat region---and in particular, on \(\mathcal{B}\)---the coordinates \emph{will not be inertial}; instead, they will \emph{rotate} (due to the orbital motion) and \emph{contract} (due to the holes' inspiral). That is, if $r$ is a coordinate radius measured from the system's center of energy, and if \(r^i\) is a radial position vector in the asymptotically flat region, then
\begin{eqnarray}\label{junkeq:outerBCshift}
\beta^i = (\mathbf{\OmegaOrbitID} \times \mathbf{r})^i + \frac{v_r}{r_o} r^i \mbox{ on } \mathcal{B}.
\end{eqnarray} Here $r_o=d_o/2$, where $d_o$ is the initial coordinate separation of the holes. The precise values of 
\(\OmegaOrbitID\) and \(v_r\) will be set so that the holes' subsequent trajectories are not eccentric (Sec.~\ref{junksec:eccentricity}).

\subsubsection{Inner boundary conditions}
\label{junksec:innerBC}
The singularities of each black hole are excised from the computational domain. The excision surface \(\mathcal{S}\) is chosen to be the apparent horizons \(\mathcal{H}\) of the two holes (labeled ``A'' and ``B''), i.e., \(\mathcal{S} = \mathcal{H}_A \bigcup \mathcal{H}_B\). This requirement leads to a boundary condition on the conformal factor (Eqs.~(28)~and~(48) of~\cite{Cook2004}):
\begin{equation}\label{junkeq:AH-BC}
\CSpatialNormal^k\partial_k \CF = 
-\frac{\CF^{-3}}{8\CLapse}\CSpatialNormal^i\CSpatialNormal^j
\left[\CLong{\Shift}_{ij}-\dtCMetric_{ij}\right]
-\frac{\CF}{4}\,\CTwoMetric^{ij}\CCD_i\CSpatialNormal_j
+\frac{1}{6}\TrExCurv\CF^3 \mbox{ on }\mathcal{S}
\end{equation}
where \(s^i\) is an outward-pointing\footnote{Here ``outward-pointing'' points away from the black hole, toward infinity.} unit normal vector on \(\mathcal{S}\), and \(h_{ij}=g_{ij}-s_i s_j = \psi^4 \left(\tilde{g}_{ij} - \tilde{s}_i \tilde{s}_j\right) = \psi^4 \tilde{h}_{ij}\) is the induced metric on \(\mathcal{S}\).

When the initial data are evolved in the comoving system, the apparent horizon (itself in equilibrium) should remain at rest. This leads to the following boundary condition on the shift (Eqs.~(36)~and~(50) of~\cite{Cook2004}):
\begin{eqnarray}
\beta^i = \alpha s^i + \Omega_r \xi^i\mbox{ on }\mathcal{S}.
\end{eqnarray} Here \(\Omega_r\) is a parameter that determines the amount of spin on the hole \emph{in addition to corotation}, and \(\xi^i\) is a conformal Killing vector within \(\mathcal{S}\). (If the holes are to have different spins, different values of \(\Omega_r\) are used on \(\mathcal{H}_A\) and \(\mathcal{H}_B\).) The $(k)^{\rm th}$ component of the quasilocal spin \cite{BrownYork,Ashtekar2003,Ashtekar-Krishnan:2004} of each hole can be written as  (e.g., Eq.~(37) of~\cite{Caudill-etal:2006})
\begin{eqnarray}\label{junkeq:BrownYorkSpin}
a^{(k)}_\mathcal{H} := \frac{1}{8\pi} \oint_{\mathcal{H}} \left(K_{ij}-g_{ij} K\right) \xi^j_{(k)} d^2 S^i,
\end{eqnarray} where \(\mathcal{H}\) is either \(\mathcal{H}_A\) or \(\mathcal{H}_B\), and \(\xi^i_{(k)}\) is a Killing vector on \(\mathcal{H}\) that defines rotations about the $k$ axis. In general, 
there are not necessarily any Killing vectors on $\mathcal{H}$; 
in such cases, the black hole spin can still be measured using 
Eq.~(\ref{junkeq:BrownYorkSpin}) but with $\xi^i_{(k)}$ taken to be 
an approximate Killing vector~\cite{Cook2007,OwenThesis}. 
In this paper, I consider only non-rotating binaries, in which \(\Omega_r\) is selected so that 
the approximate-Killing-vector spin \(a^{(k)}_\mathcal{H}\) is very close to zero. (The specific method that I use to measure the black-hole spins is described in Appendix A of Ref.~\cite{Lovelace2008}.)

The inner condition on \(\alpha\) is a gauge choice~\cite{Cook2004}; i.e., it does not affect the physical content of the initial data. The particular choices used in this paper are discussed in Sec.~\ref{junksec:SBSMetric}.

\section{Nonspinning, non-eccentric binary-black-hole initial data}
\label{junksec:SBSMetric}
\subsection{Conformally flat data}
Initial data for binary-black-holes are typically assumed to be conformally flat:
\begin{equation}
\tilde{g}_{ij} = f_{ij}.
\end{equation} In Sec.~\ref{junksec:datasets}, the conformally flat, maximally sliced data set CFMS (which is ``30c'' in Ref.~\cite{boyleEtAl2007}) chooses
\begin{eqnarray}
K = 0,\\
\partial_r (\alpha\psi) = 0 \mbox{ on }\mathcal{S}.
\end{eqnarray}

\subsection{Conformally curved data}
To reduce the amount of junk radiation, I adopt a curved conformal metric. In this section, I build up a suitable conformal metric for two non-spinning black holes in an initially circular orbit.

\subsubsection{Schwarzschild with maximal slicing}

The Schwarzschild metric can be split into maximal slices (i.e., slices with \(K = 0\).) With maximal slicing, the Schwarzschild spacetime is (e.g., Eq. (52) of \cite{Cook2004}):
\begin{eqnarray}\label{junkeq:MSSNotCF}
g_{ij}^o dx^i dx^j & = & \alpha^{-2} dR^2 + R^2 d\Omega^2,\nonumber\\
\alpha & = & \sqrt{1-\frac{2 M_S}{R}+\frac{C^2}{R^4}},\nonumber\\
\beta^i & = & \frac{C}{R^2}\alpha e_R^i,
\end{eqnarray} where \(R\) is the Schwarzschild areal radial coordinate and \(e_R^i\) is a unit vector normal to constant-\(R\) surfaces. The choice of \(C\) specifies which maximal slicing is used (and thence the coordinate radius of the horizon). To facilitate comparison with~\cite{EccentricityPaper}, in this paper I choose \(C=1.737\), which implies the horizon radius is at \(r_{exc}=0.8595\), and $M_S=1$. 

\subsubsection{Metric of a boosted Schwarzschild hole}

To construct the metric of a boosted, Schwarzschild hole, I begin with the Schwarzschild metric \(g_{\mu\nu}^{o}\) [Eq. (\ref{junkeq:MSSNotCF})]. Then, I apply the following two coordinate transformations:
\begin{enumerate}
\item First, I apply a radial coordinate transformation to make the spatial metric conformally flat. \emph{Any} spherically-symmetric metric can be made conformally flat by i) writing it as
\begin{eqnarray}
g_{ij} dx^i dx^j = f(R) dR^2 + R^2 d\Omega^2,
\end{eqnarray} where \(R\) is an areal radial coordinate, and then ii) making a radial coordinate transformation \(r=r(R)\) so that the metric is conformally flat:
\begin{eqnarray}
f(R) dR^2 + R^2 d\Omega^2 = \psi^4\left(dr^2 + r^2 d\Omega^2\right)\\
\Rightarrow \psi^2 = R/r \mbox{ and } \frac{dr}{dR} = \frac{r}{R}\sqrt{f(R)}. \label{junkeq:ODE}
\end{eqnarray} For a given \(f(R)\), determining \(r(R)\) amounts to solving a first-order ordinary differential equation. For the choice $f(R)=\alpha^{-2}$ [cf. Eq.~(\ref{junkeq:MSSNotCF})], the analytic solution to the ODE is unknown, so the ODE (\ref{junkeq:ODE}) is solved numerically.
\item Once the spatial metric is conformally flat, I give the Schwarzschild hole a velocity \(v\) in the \(y\) direction by performing a Lorentz boost. The resulting spacetime metric is \(g_{\mu\nu}^{\rm Boost}(v)\).
\end{enumerate}

\subsubsection{Superposing two boosted, non-spinning holes}
Suppose two nonspinning black holes, ``A'' and ``B,'' are initially separated by coordinate distance \(d_o\) along the x axis. Suppose they are in a circular orbit about the z axis with speed \(v=\OmegaOrbitID d_o/2\). 

I seek free data that accurately describes the boosted black holes. The simplest choice is to merely superpose two boosted-Schwarzschild black-holes:
\begin{eqnarray}\label{junkeq:NBSS}
\tilde{g}_{ij} & = & f_{ij}+\left(g_{ij}^A - f_{ij}\right) + \left(g_{ij}^B - f_{ij}\right),\\
K & = & K^A + K^B.
\end{eqnarray} For hole A, \(g_{ij}^A\) and \(K^A\) are obtained by translating \(g_{\mu\nu}^{\rm Boost}(v)\) so that the hole is centered about \((x,y,z)=(d_o/2,0,0)\). Likewise, for hole B, \(g_{ij}^B\) and \(K^B\) are obtained by translating \(g_{\mu\nu}^{\rm Boost}(-v)\) so that the hole is centered about \((x,y,z)=(-d_o/2,0,0)\). Note that as $v\to0$, $\tilde{g}_{ij}\to\psi_o^4 f_{ij}$ and $K\to 0$; after absorbing $\psi_o$ into the conformal factor $\psi$, the $v=0$ superposed-boosted-Schwarzschild data in fact reduces to the well-examined (conformally flat) quasiequilibrium data by Cook and Pfeiffer~\cite{Cook2004}.
\begin{figure}
\vspace{1mm}
\centerline{\includegraphics[width=4.5in]{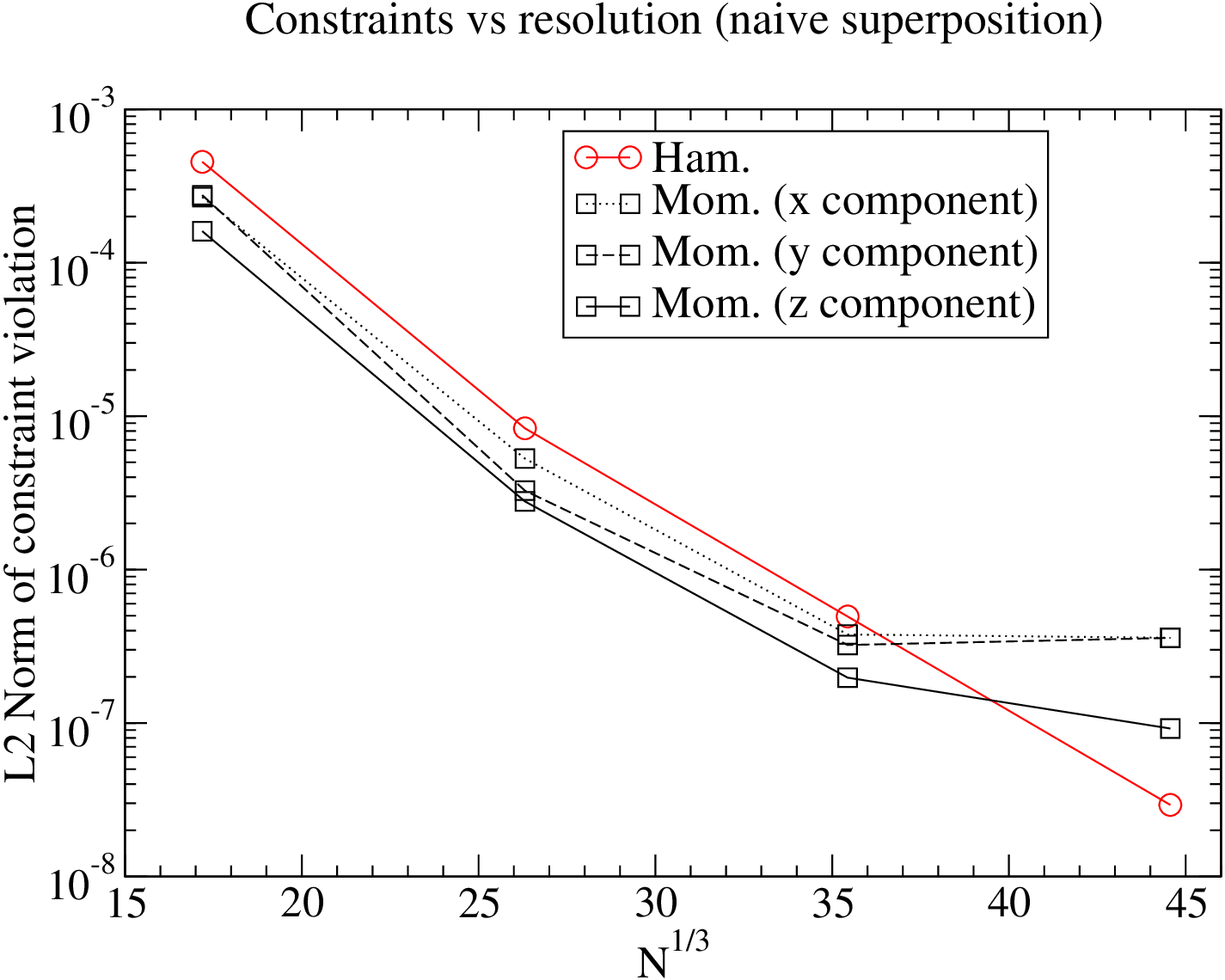}}
\caption{\emph{Color online.} When the conformal metric $\tilde{g}_{ij}$ is given by Eq.~(\ref{junkeq:NBSS}), the constraints do not decrease exponentially with increasing resolution (indicated here by the number of gridpoints $N$ on the computational grid).\label{fig:JunkFig1}}
\end{figure}

I used the Caltech-Cornell pseudospectral code~\cite{Pfeiffer2003} to solve the XCTS equations with the free data and boundary conditions described previously. When the conformal metric $\tilde{g}_{ij}$ is given by Eq.~(\ref{junkeq:NBSS}), I find that the elliptic solver does not converge [Fig.~\ref{fig:JunkFig1}].

\begin{figure}
\vspace{1mm}
\centerline{\includegraphics[width=4.5in]{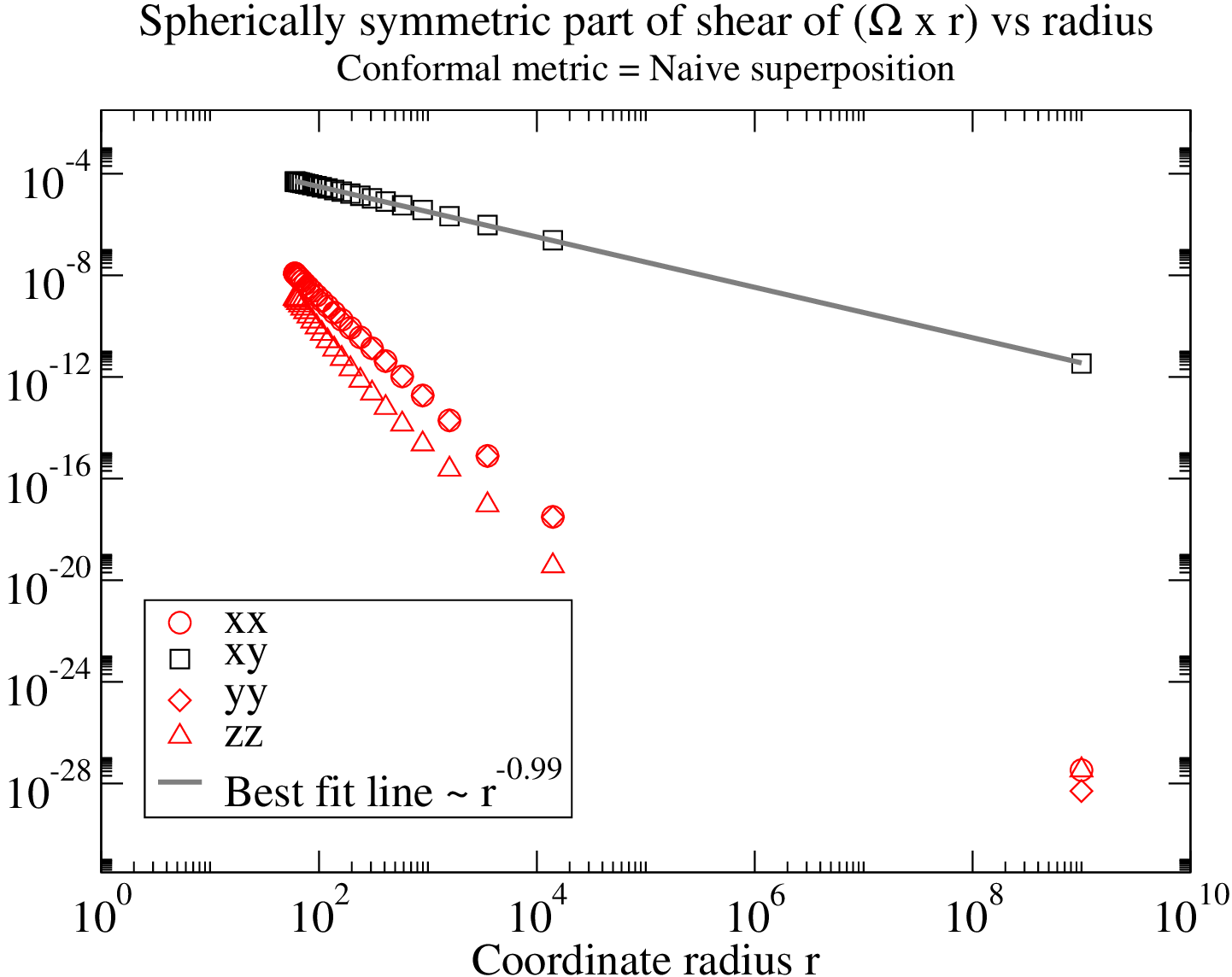}}
\caption{\emph{Color online.} Magnitude of selected components of $\tilde{[\mathbb{L}} \left(\mathbf{\OmegaOrbitID} \times \mathbf{r}\right)]^{ij}$; the plot shows the absolute value of the angular average as a function of radius for each Cartesian component. The component $[\tilde{\mathbb L}\left(\mathbf{\OmegaOrbitID}\times \mathbf{r}\right)]^{xy}$ includes a spherically-symmetric term that decays only as $1/r$ when the conformal metric is given by Eq.~(\ref{junkeq:NBSS}); this term causes the conformal factor to diverge logarithmically as $r\rightarrow\infty$. The other nonzero components decay as $1/r^2$ or faster. \label{fig:JunkFig3}}
\end{figure}

The source of the difficulty can be seen by inserting the outer boundary conditions (\ref{junkeq:outerBCpsi}), (\ref{junkeq:outerBClapsePsi}), and (\ref{junkeq:outerBCshift}) into the Hamiltonian constraint [the first of Eq.~(\ref{junkeq:constraints})], which gives an equation of the form
\begin{eqnarray}
-\tilde{\nabla}^2 \psi - [\tilde{\mathbb L}\left(\mathbf{\OmegaOrbitID}\times \mathbf{r}\right)]^{ij} [\tilde{\mathbb L}\left(\mathbf{\OmegaOrbitID}\times \mathbf{r}\right)]_{ij}+ \cdots = 0.
\end{eqnarray} The source term shown here vanishes when the metric is conformally flat, because \((\mathbf{\OmegaOrbitID}\times \mathbf{r})^i\) is  a conformal Killing vector of flat space. But when the conformal metric is the naive superposition (\ref{junkeq:NBSS}), the term \([\tilde{\mathbb L}\left(\mathbf{\OmegaOrbitID}\times \mathbf{r}\right)]^{xy}\) contains a spherically-symmetric part that decays only as \(O(1/r)\) [Fig. \ref{fig:JunkFig3}]. It follows that \(\psi \rightarrow {\rm const} \times \log r\) as \(r\rightarrow\infty\), but this is incompatible with the requirement that \(\psi\rightarrow 1\) as \(r\rightarrow\infty\).

\subsubsection{Scaling the non-flat terms by Gaussians}
\begin{figure}
\vspace{1mm}
\centerline{\includegraphics[width=4.5in]{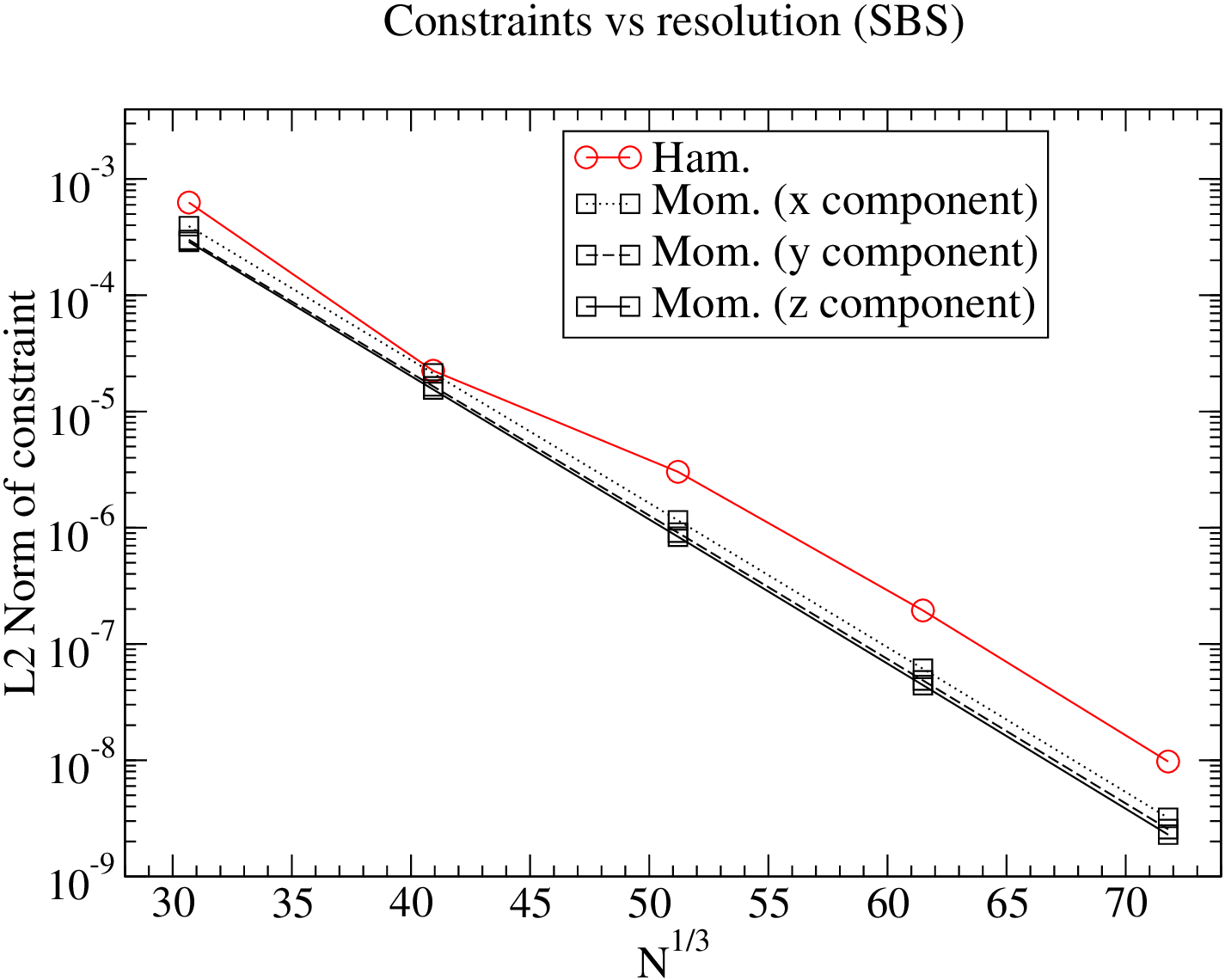}}
\caption{\emph{Color online.} Constraints as a function of resolution (indicated by the number of gridpoints $N$) when the conformal metric is $g_{ij}^{\rm SBS}$, which is curved near the black holes but flat far away. The constraints decrease exponentially with increasing resolution, as expected. \label{fig:JunkFig4}}
\end{figure}
The inconsistency described in the previous subsection can be avoided by requiring that the non-flat terms fall off sufficiently quickly far from the holes. This can be accomplished by scaling the non-flat terms by Gaussians:
\begin{eqnarray}
\tilde{g}_{ij} & = & \tilde{g}_{ij}^{\rm SBS} := f_{ij}+e^{-r_A^2/w^2}\left(g_{ij}^A - f_{ij}\right) + e^{-r_B^2/w^2}\left(g_{ij}^B - f_{ij}\right),\label{junkeq:SBSMetric}\\
K & = & K^{\rm SBS} := e^{-r_A^2/w^2}K^A + e^{-r_B^2/w^2}K^B.
\end{eqnarray} Here ``SBS'' stands for ``superposed, boosted Schwarzschild,'' and \(r_A\) and \(r_B\) are the coordinate distances from the centers of holes A and B, respectively. The width \(w\) can be adjusted to minimize the junk radiation; further investigation is needed to find an optimal choice for \(w\). The Gaussians used here may seem similar to the attenuation functions of~\cite{MM2000}; however, the attenuation functions here approach zero at large radii, whereas those of~\cite{MM2000} approach unity. In Sec.~\ref{junksec:datasets}, data set SBS uses \(w=20 r_{exc}\). With this choice, the elliptic solver converges, and the constraints decay exponentially with resolution, as expected [Fig. \ref{fig:JunkFig4}]. 

Finally, as noted in Sec.~\ref{junksec:innerBC}, the inner boundary condition on the lapse $\alpha$ is a gauge choice; in set SBS, the condition is
\begin{eqnarray}
\alpha \psi = 1+e^{-r_A^2/w^2}\left(\alpha^A-1\right)+e^{-r_B^2/w^2}\left(\alpha^B-1\right)\mbox{ on }\mathcal{S}.
\end{eqnarray} Here $\alpha^A$ is obtained by translating \(g_{\mu\nu}^{\rm Boost}(v)\) so that hole A is centered about \((x,y,z)=(d_o/2,0,0)\); likewise, $\alpha^B$ is obtained by translating \(g_{\mu\nu}^{\rm Boost}(-v)\) so that hole B is centered about \((x,y,z)=(-d_o/2,0,0)\).

\section{Comparing the junk radiation of conformally flat and superposed-boosted-Schwarzschild initial data}
\subsection{Initial data sets}
\label{junksec:datasets}
\begin{table}
\centerline{\begin{tabular}{c|cc}
\hline\hline\\
{\bf Quantity} & {\bf CFMS} & {\bf SBS}\\\hline
$M \OmegaOrbitID$ & 0.016708 & 0.016405  \\ 
$v_r$ & $-4.3 \times 10^{-4}$ & $-3.5\times 10^{-4}$ \\ 
\hline
$M_{\rm irr}$ & 1.039 & 1.027\\ 
$M :=2M_{\rm irr}$ & 2.078 & 2.044\\ 
$M_{\rm ADM}$ & 2.062 & 2.039\\\hline 
$|a^x/M_{\rm irr}^2|$ and $|a^y/M_{\rm irr}^2|$ & $< 2\times 10^{-15}$ & $< 2\times 10^{-10}$\\ 
$|a^z/M_{\rm irr}^2|$ & $< 6 \times 10^{-8}$ & $< 5 \times 10^{-6}$
\\\hline
$d_o / M$ & 14.44 & 14.60\\ 
$s_o / M$ & 17.37 & 17.51\\ 
\hline\hline 
\end{tabular}}
\caption{\label{junktab:datasets} A comparison of the two initial data sets presented in this paper. Set CFMS is 
the conformally flat initial data set ``30c'' in 
Ref.~\cite{boyleEtAl2007}, and set SBS uses a conformal metric that is a superposition of two boosted Schwarzschild black holes. The initial data sets describe physically comparable situations: the masses and separations agree to within about 1\%, and the frequencies agree to within about 2\%. The radial velocities are comparable [and are chosen so that the eccentricity is small (Fig.~\ref{fig:JunkFig5})], and the spins of the holes are close to zero in both cases.}
\end{table}  

In this section, I compare the amount of junk radiation during
evolutions of 
two initial data sets, both of which correspond to equal-mass, 
non-spinning binary black holes 
in nearly-circular obits; the holes are about 15 orbits away from 
merger. The data sets are 
i) a conformally flat, maximally sliced set (CFMS) which is identical 
to the initial data set ``30c'' that is constructed 
and evolved in Ref.~\cite{boyleEtAl2007}, and ii) a superposed-boosted-Schwarzschild set (SBS) which is constructed using 
the method described in Sec.~\ref{junksec:SBSMetric}. 

Table~\ref{junktab:datasets} compares some physical properties of the two data sets. The table lists three measures 
of mass: i) the irreducible mass
$M_{\rm irr} = \sqrt{A_{\rm AH}/16\pi}$, where \(A_{\rm AH}\) is the 
area of one hole's apparent horizon, ii) the sum $M$ 
of the holes' irreducible masses, and iii) the Arnowitt-Deser-Misner 
(ADM) Mass, a measure of the total energy in the system, defined by the following surface integral at spatial infinity:
\begin{eqnarray}
M_{\rm ADM} = \frac{1}{16\pi}\int_\infty \left(\partial_j g_{ij} - \partial_{i} g_{jj}\right) d^2 S^i.
\end{eqnarray} In practice, the integral for $M_{\rm ADM}$ 
is evaluated on the outer boundary of the initial-data grid (i.e., on \(\mathcal{B}\), which is a sphere whose radius is approximately \(10^9 M_{\rm irr}\)). 
As shown in Table~\ref{junktab:datasets}, both $M_{\rm irr}$  and 
$M_{\rm ADM}$ agree within about $1\%$.
Note that in the rest of this section, 
following Ref.~\cite{boyleEtAl2007}, I will typically
express quantities with dimension in terms of the mass 
$M$.

The spins of the holes are computed using the quasilocal 
approximate-Killing-vector spin [Eq.~(\ref{junkeq:BrownYorkSpin}) 
and the surrounding discussion]. 
Table~\ref{junktab:datasets} list the quasilocal spin for both 
the CFMS and SBS data sets; in both cases, the spin is close to zero.

The initial coordinate separation of the holes \(d_o\) is given 
for sets CFMS and SBS in Table~\ref{junktab:datasets}. Also given 
is the initial ``proper separation'' \(s_o\), which is defined in terms of the following line integral along the x-axis of the 
comoving frame:
\begin{equation}
s_o = \int ds = \int_{-d/2+r_{exc}}^{d/2-r_{exc}} dx \sqrt{g_{xx}}
\end{equation} Here the limits of integration are the coordinate locations where the holes' apparent horizons intersect the x axis, 
and the radius $r_{exc}$ is the coordinate 
radius of the apparent horizon. Note that the ``proper separation'' is 
coordinate-dependent (since the integral is taken along the x-axis, 
rather than along a geodesic of extremal length) and slicing-dependent 
(since the distance is measured within the spatial slice).

\label{junksec:eccentricity}
\begin{figure}
\vspace{1mm}
\centerline{\includegraphics[width=4.5 in]{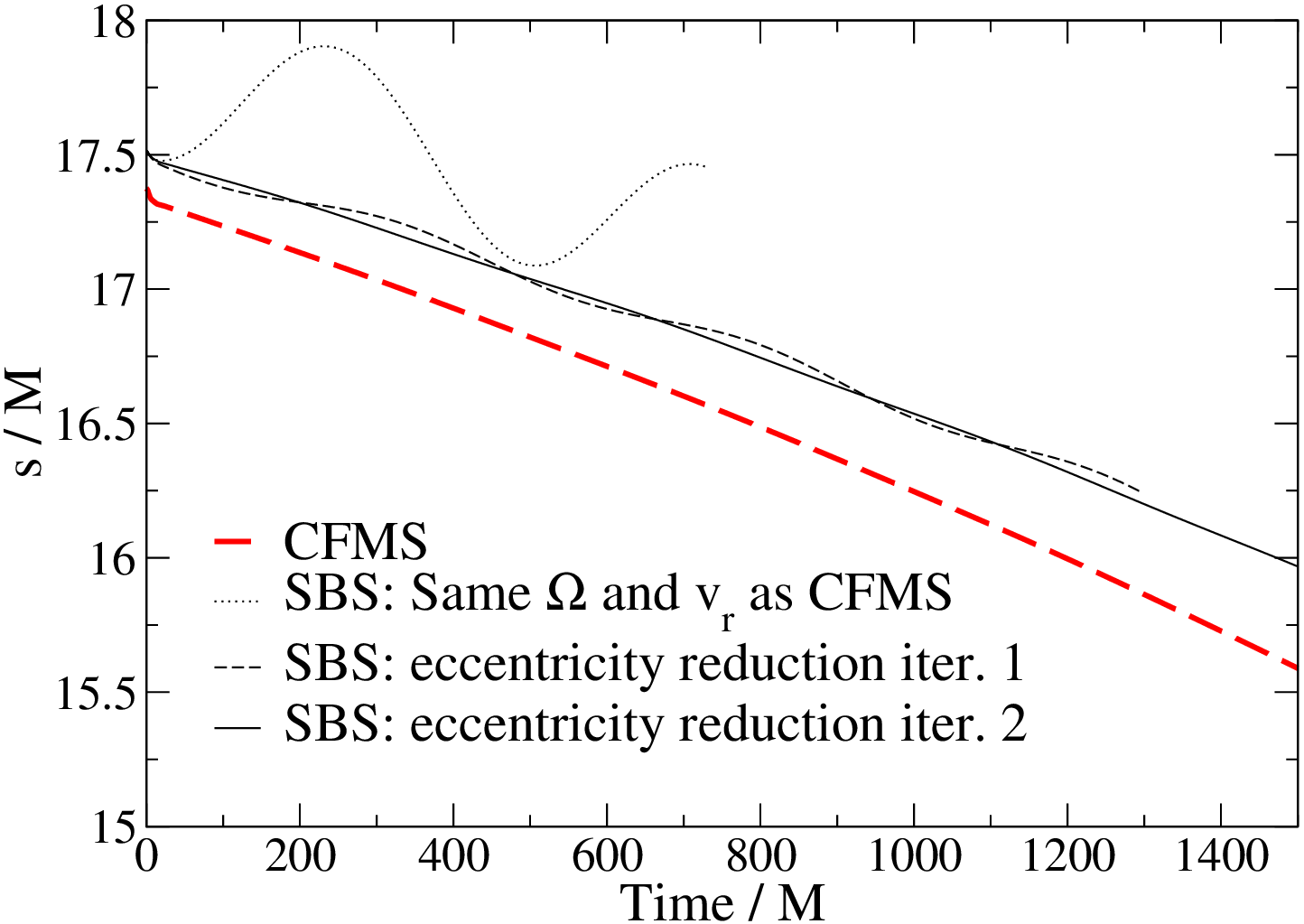}}
\caption{\emph{Color online.} Reducing the eccentricity in the SBS initial data set. The proper separation $s$ vs. time is shown for the conformally flat, maximally sliced (CFMS) data set (dashed red line). The corresponding SBS data set with the same choices for $\OmegaOrbitID$ and $v_r$ is much more eccentric (dotted black line), but two iterations of the algorithm described in Sec.~4 of 
Ref.~\cite{EccentricityPaper} greatly reduce the orbital eccentricity (dashed and solid black lines). All evolutions shown in this plot were 
performed at resolutions of $46.8^3$ gridpoints. 
\label{fig:JunkFig5}}
\end{figure}

The initial orbital frequency $\OmegaOrbitID$ and radial velocity $v_r$ 
listed in Table~\ref{junktab:datasets} are chosen to reduce the 
holes' orbital eccentricity. The eccentricity reduction for set CFMS is described in Sec. IIA of Ref.~\cite{boyleEtAl2007}. To reduce the eccentricity of the SBS data, I initially guess that \(\OmegaOrbitID\) and \(v_r\) have the same coordinate values as the CFMS values; then, I tune them using the iterative scheme described in Sec.~4 of~\cite{EccentricityPaper}. Each iteration would completely remove the eccentricity if the binary were Newtonian; in the relativistic case, successive iterations converge to non-eccentric orbits. Figure~\ref{fig:JunkFig5} illustrates the eccentricity reduction by 
showing the proper separation as a function of time for set CFMS and 
also for several iterations of SBS data. The set ``SBS: eccentricity reduction iter. 2'' is the set SBS described in 
Table~\ref{junktab:datasets} and used in the remainder of this 
paper.

\subsection{Evolutions}\label{sec:Evolutions}
The CFMS and SBS data sets were evolved using the Caltech-Cornell pseudospectral evolution code ${\tt SpEC}$~\cite{Scheel2006}. 
I used the same evolution methods, equations, and boundary conditions 
as those described in Ref.~\cite{boyleEtAl2007}. In particular, 
I use the first-order generalized-harmonic system with constraint damping that is derived in Ref.~\cite{Lindblom2006}. The outer boundary conditions, derived in Ref.~\cite{Lindblom2006} and augmented with improved conditions on the gauge fields in Refs.~\cite{Rinne2006,Rinne2007}, preserve the constraints and enforce a ``no-incoming-radiation'' requirement by freezing the Newman-Penrose scalar $\Psi_0$.

The evolution grid's outer boundary is at approximately \(900 M_{\rm irr}\). The excision boundaries are slightly inside the apparent horizons; this is accomplished by extrapolating the initial data from \(\mathcal{S}\) to points slightly inside \(S\). (For set CFMS, the evolution-grid excision spheres are at radius \(r_{ev}=0.97 r_{exc}\); to accommodate nonspherical horizons, the evolution-grid excision spheres are at \(r_{ev}=0.93 r_{exc}\) for set SBS.) Aside from differences in extrapolation, the evolutions of sets CFMS and SBS 
used identical computational domains and identical numerical  resolutions. 

\begin{figure}
\vspace{1mm}
\centerline{\includegraphics[width=6 in]{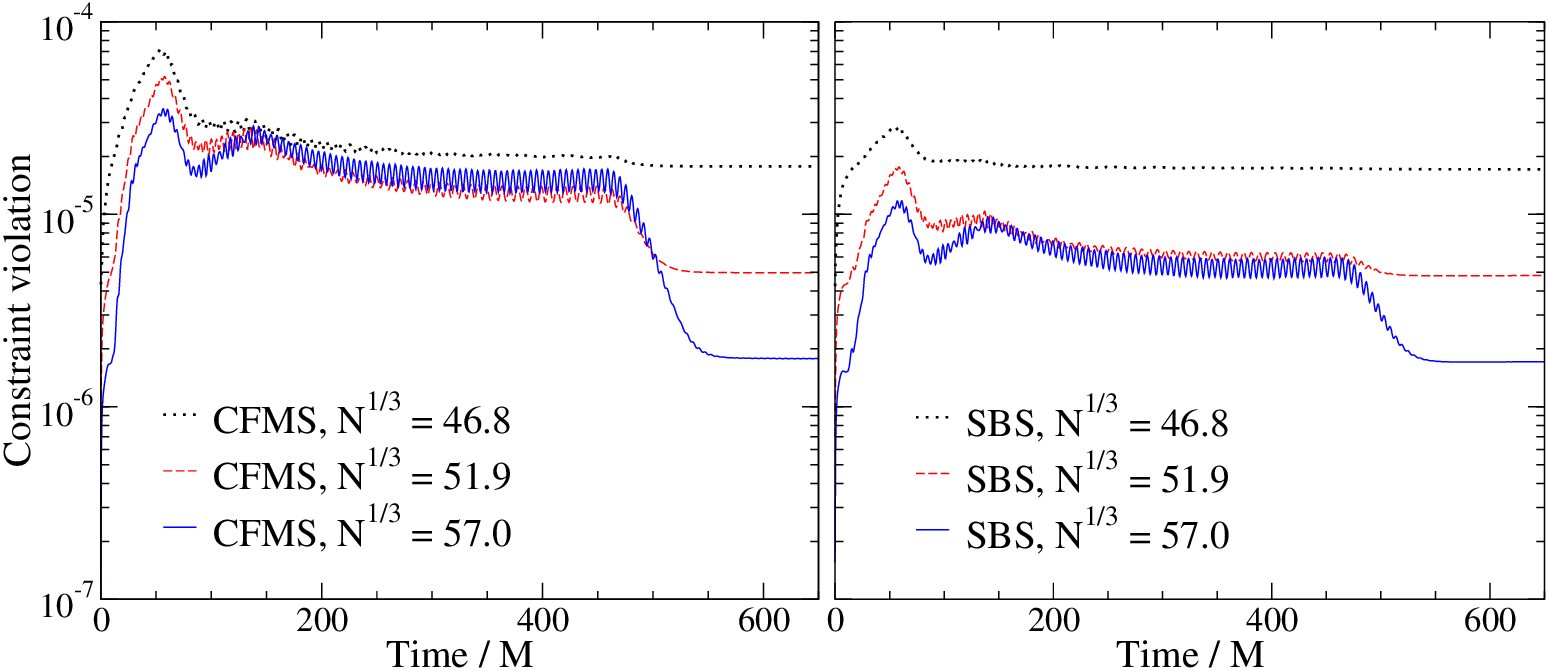}}
\caption{{\it Color online.} The constraint violation plotted as 
a function of time for evolutions of 
conformally flat, maximally sliced (CFMS, left panel) and 
superposed-boosted-Schwarzschild (SBS, right panel) initial data. 
The constraints are shown for three different spatial resolutions 
(labeled by their corresponding number of gridpoints $N$).
\label{fig:JunkFigGhCe}}
\end{figure}

Each evolution of sets CFMS and SBS were performed at 
three different numerical resolutions, 
corresponding to $46.8^3$, $51.9^3$, and $57.0^3$ gridpoints, 
respectively. 
Figure~\ref{fig:JunkFigGhCe} compares the constraint 
violations (which are computed in the same way as in footnote 8 of Ref.~\cite{Lovelace2008}) at each resolution. At late times 
($\gtrsim 500 M$), after the initial burst of junk radiation has 
left the computational domain, the constraints converge exponentially, as expected---but while the junk radiation is on the grid, the constraints do not appear to decrease significantly between the $N=51.9^3$ and $N=57.0^3$ resolutions. This poor convergence is not surprising; because junk radiation typically has a much higher frequency than the physical 
gravitational waves, the spurious radiation requires much higher numerical resolutions to be accurately resolved. In practice, such high resolutions are never used, 
because the added computational cost (which would likely 
be prohibitively expensive) is not necessary to 
adequately resolve the astrophysically-realistic content of the 
simulation. 

Because my motivation is to reduce the observed amount 
of spurious radiation, not to study its physical content, 
in this paper I do not attempt to better resolve the junk radiation. Instead, in the next subsection,
I simply compare the amounts of junk radiation that can be seen 
for the three 
resolutions shown in Fig.~\ref{fig:JunkFigGhCe}. 

\subsection{Junk radiation comparison}
\label{junksec:junkcompare}
\begin{figure}
\vspace{1mm}
\centerline{\includegraphics[width=6 in]{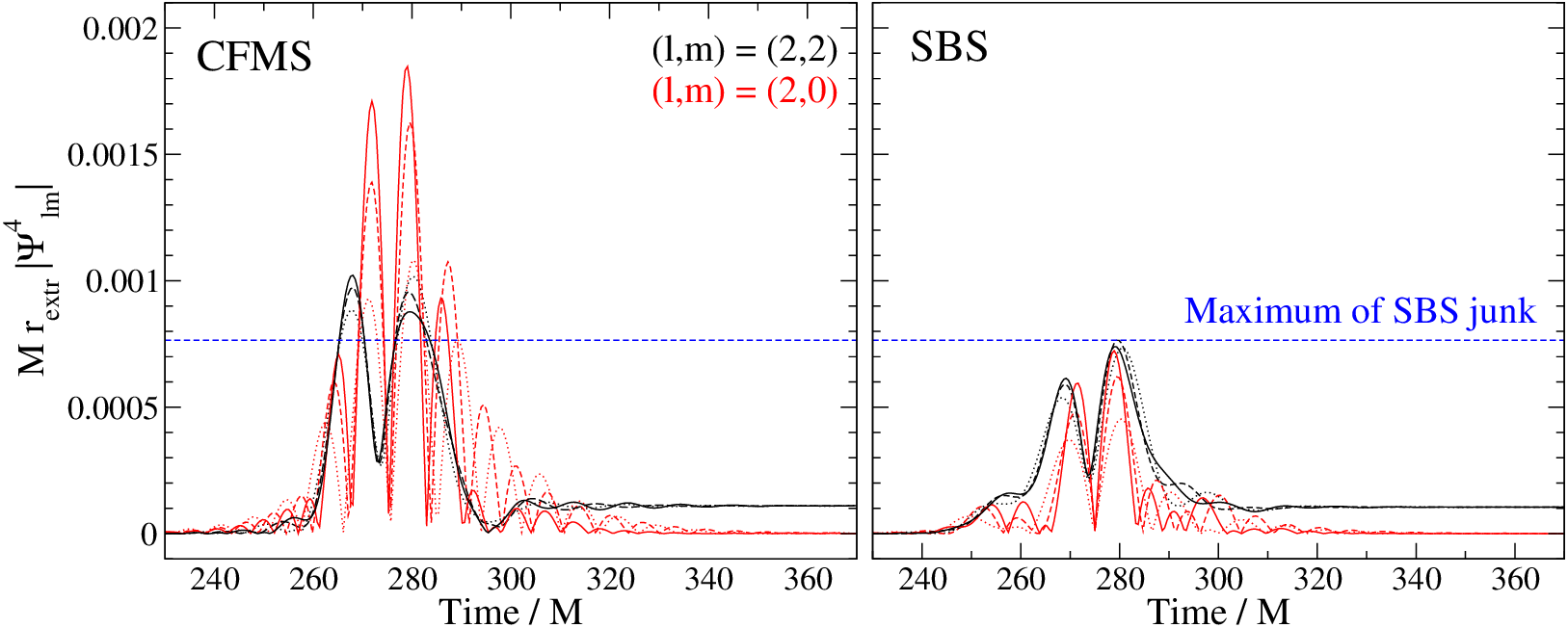}}
\caption{{\it Color online.} A comparison of the dominant (quadrupolar) modes of the junk radiation for conformally flat (CFMS, 
left panel) and superposed-boosted-Schwarzschild 
(SBS, right panel) initial data. The 
$(\ell,m)=(2,0)$ mode (red) and 
$(\ell,m)=(2,2)$ mode (black) of $\left|\Psi_4\right|$ are shown at 
at three different spatial resolutions: $46.8^3$ (dotted lines), 
$51.9^3$ (dashed lines), and $57.0^3$ (solid lines). 
The waves are extracted on a coordinate sphere of radius 
$265 M$. The blue, dashed horizontal line is the maximum amplitude of the SBS junk.
\label{fig:JunkFig6}}
\end{figure}

\begin{figure}
\vspace{1mm}
\centerline{\includegraphics[width=6 in]{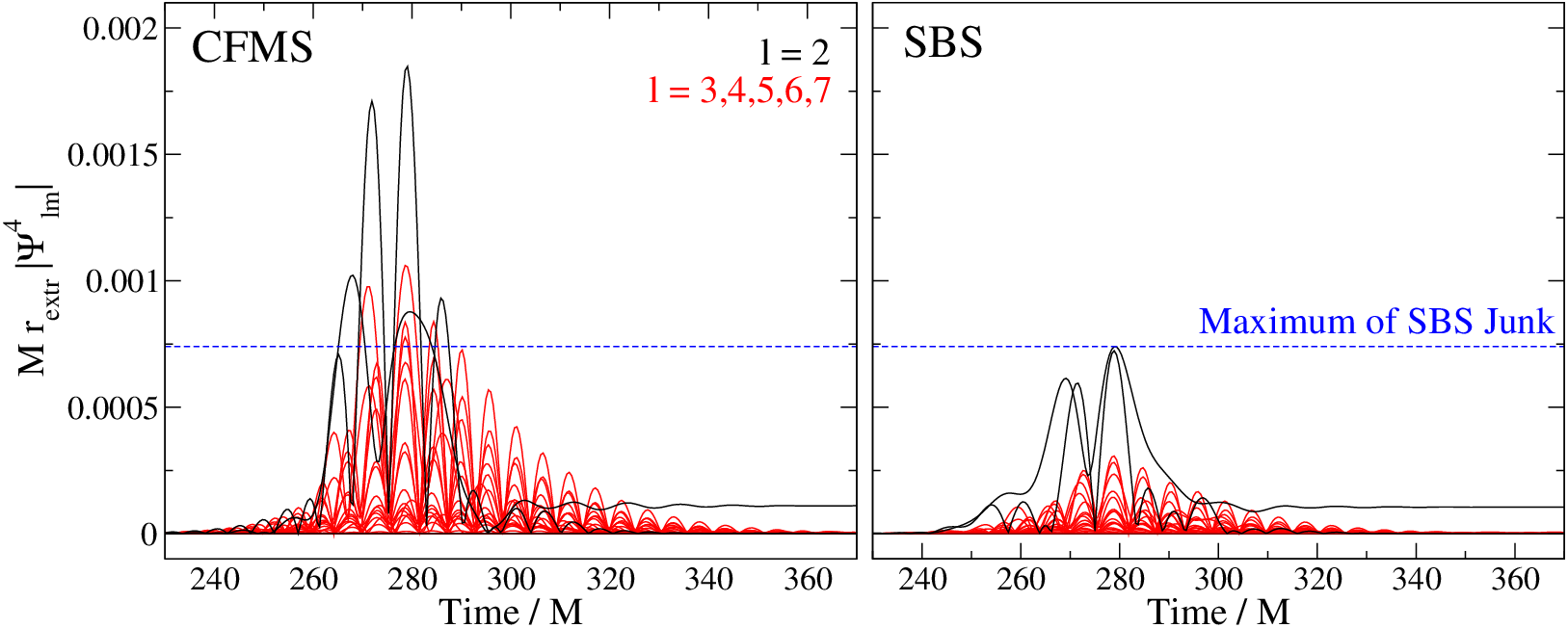}}
\caption{\emph{Color online.} A comparison of the gravitational waves extracted at coordinate radius $r_{\rm extr}=265 M$ 
for the conformally flat, maximally sliced (CFMS, left panel) and superposed-boosted-Schwarzschild (SBS, right panel) initial data. The $\ell=2$ spherical-harmonic 
modes are shown in black, and higher modes are 
shown in red. All of the modes of $\left|\Psi_4\right|$ shown here 
were computed with a spatial resolution of $N=57.0^3$ gridpoints. 
The blue, dashed, horizontal line is the maximum amplitude of the SBS junk.
\label{fig:JunkFig7}}
\end{figure}

The gravitational waves are extracted from the CFMS and SBS evolutions at the same coordinate radius 
\(r_{\rm extr}\). Specifically, the simulation computes the Newman-Penrose scalar \(\Psi_4\), which at large radii is related to the outgoing +-polarized and $\times$-polarized gravitational-wave amplitudes by
\begin{equation}
\Psi_4 = \frac{d^2}{dt^2} h_+ - i \frac{d^2}{dt^2} h_\times.
\end{equation} The scalar \(\Psi_4\) is evaluated on a sphere of radius \(r_{\rm extr}= 265 M \) and then expanded in spin-weighted-spherical-harmonic modes $\Psi^4_{\ell m}$. 
(For further details on the wave-extraction method 
used here, see Sec.~5.3 of Ref.~\cite{EccentricityPaper}.) 

At early times, the waveform consists of spurious gravitational waves; they are recognizable as such by their frequencies, which are much higher than the dominant frequencies $f\sim\OmegaOrbitID/\pi$ of the physical, quadrupolar gravitational waves. Figure~\ref{fig:JunkFig6} plots dominant, quadrupolar components of 
$M r_{\rm extr}\left|\Psi^4\right|$; they have a frequency $f\sim (15 M)^{-1}$, which is significantly higher than the physical frequency $f\sim\OmegaOrbitID/\pi\sim(200 M)^{-1}$. Figure~\ref{fig:JunkFig7} compares all of the spherical harmonic modes of 
$M r_{\rm extr}\left|\Psi^4\right|$ from \(\ell=2\) through \(\ell=7\) with $m\ge 0$. 

Following the 
initial high-frequency spurious radiation, the astrophysical waves 
are dominated by the $(\ell,m)=(2,2)$ mode, with 
$M r_{\rm extr}\left|\Psi^4_{22}\right|$ a 
dimensionless slowly-growing amplitude that is essentially constant during the time shown in 
Fig.~\ref{fig:JunkFig7}; during this interval, this amplitude is 
significantly smaller than the peaks of the junk radiation.

\begin{figure}
\vspace{1mm}
\centerline{\includegraphics[width=6 in]{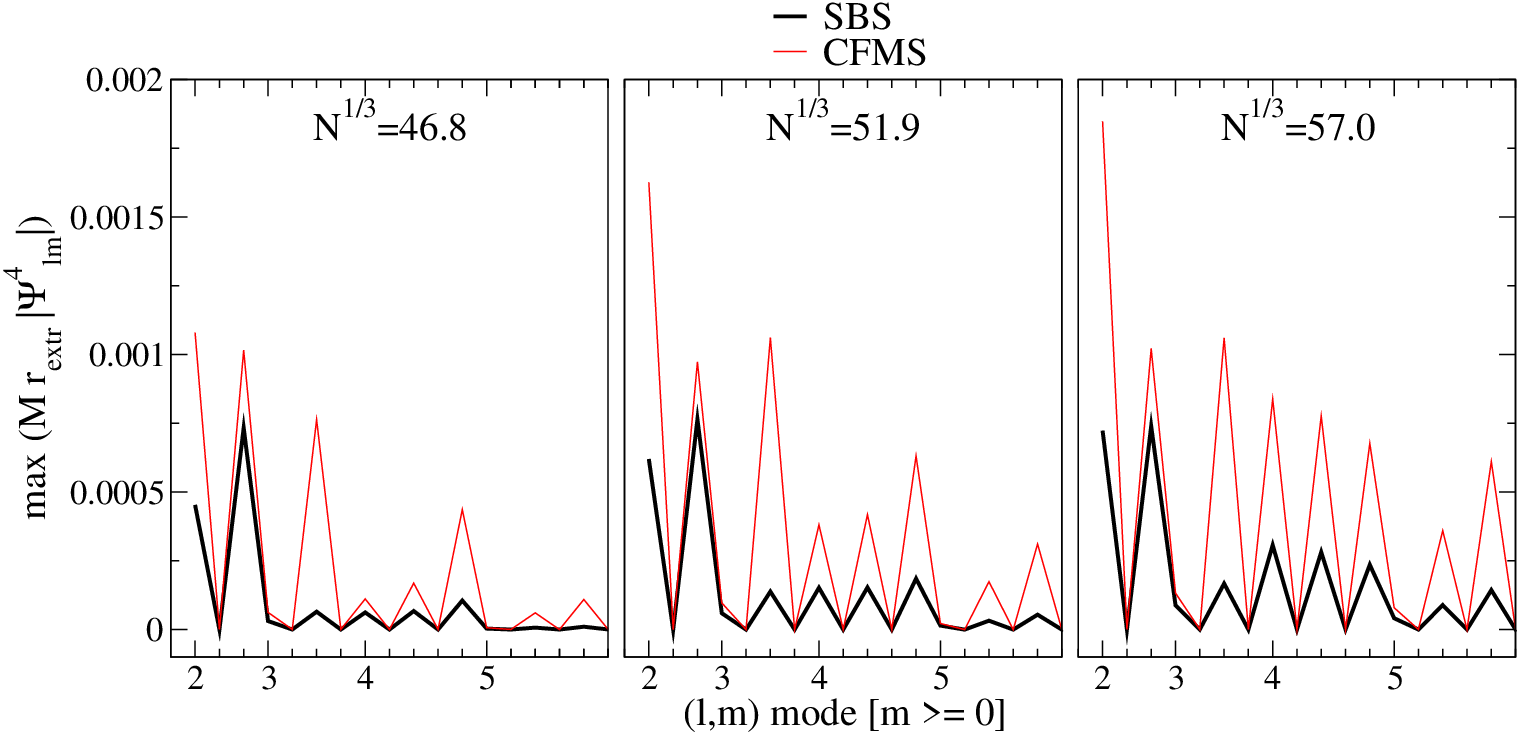}}
\caption{\emph{Color online.} A comparison of the maximum amplitude of 
the spherical harmonic modes of $\left|\Psi^4_{\ell m}\right|$ during 
the time interval shown in Figs.~\ref{fig:JunkFig6}--\ref{fig:JunkFig7}. The $(\ell,m=0)$ modes are labeled by $\ell$, and the $(\ell,m>0)$ modes are indicated by small tick marks (with $m$ increasing from left to right). 
The three panels correspond to three different spatial resolutions, with 
resolution increasing from left to right.
The waves are extracted at a radius of $r_{\rm extr}=265 M$. 
At each resolution, the nonvanishing conformally flat, maximally sliced (CFMS, red, thin line) 
modes are larger than the 
corresponding superposed-boosted-Schwarzschild (SBS, thick, black line) mode. Only the modes up through $\ell=5$ with 
$m\ge 0$ are shown.
\label{fig:JunkFigMaxAbs}}
\end{figure} 

As a simple measure of the amount of junk radiation present in each 
$(\ell,m)$ mode, Fig.~\ref{fig:JunkFigMaxAbs} shows the maximum 
value of $M r_{\rm extr} \left|\Psi^4_{\ell m}\right|$ 
for the modes from $\ell=2$ through $\ell=5$. Only modes with 
$m\ge 0$ are shown. The numerical 
values of the maxima vary with increasing resolution; this 
is another indication that the junk radiation is underresolved. However, 
for each resolution, the SBS maximum is smaller than the corresponding 
CFMS maximum. For most of the modes with large contributions to the junk radiation---including the $(\ell,m)=(2,0)$ mode, which is the 
leading contributor to the junk---the improvement is about a factor of 
two or more; however, for the $(\ell,m)=(2,2)$ mode, the 
improvement is much more modest: the SBS maximum is about 
75\% of the corresponding CFMS maximum. The smaller improvement might 
be because the SBS initial data makes no attempt to model the 
initial gravitational-wave content; if so, then the 
spurious gravitational radiation might be able to be reduced even 
more by adding post-Newtonian terms to 
$\tilde{g}_{ij}$ and $K$ (such as, e.g., terms based on the free data used in Ref.~\cite{Kellyetal2007}).

\section{Conclusion}
\label{junksec:conclusion}
The junk radiation in binary-black-hole simulations can be significantly reduced by using superposed-boosted-Schwarzschild initial data instead of conformally flat data. For the case of two non-spinning black holes initially 15 orbits from merger, the amplitude of the junk gravitational waves decrease, 
with most spherical-harmonic modes, including the leading contributors to the junk radiation, decreasing by a factor of order 2 
or more. However, a significant amount of junk radiation is still present in evolutions of the superposed-boosted-Schwarzschild data, which also does not attempt to account for the initial gravitational-wave content. Besides attempting to incorporate gravitational waves into the initial data, future studies could investigate the amount of spurious radiation in the superposed-Kerr-Schild data in Ref.~\cite{Lovelace2008}, since one might speculate that lower amounts of spurious radiation play a role in letting the superposed-Kerr-Schild data lead to evolutions with larger spins than are possible with Bowen-York initial data.

\begin{ack}
I am pleased to acknowledge Lee Lindblom for 
many helpful discussions and 
for suggesting the project described in this paper, 
Harald Pfeiffer for many helpful discussions and comments on this 
manuscript, and   
Lawrence Kidder, Niall \'O Murchadha, Robert Owen, Mark Scheel, and 
Kip Thorne for 
helpful discussions. The numerical simulations presented in this 
paper were done with the Spectral Einstein Code ({\tt SpEC}) developed 
principally by Lawrence Kidder, Harald Pfeiffer, and Mark Scheel. 
The numerical solution of equation~(\ref{junkeq:ODE}) was obtained using code written by Gregory Cook. 
This work was supported in part by grants from the Sherman 
Fairchild Foundation to Caltech and Cornell and from the 
Brinson Foundation to Caltech; by NSF Grants No. PHY-0652952, 
No. DMS-0553677, and No. PHY-0652929 and NASA Grant No. NNG05GG51G 
at Cornell; and by NSF Grants No. PHY-0601459, No. PHY-0652995, and 
No. DMS-0553302 and NASA Grant No. NNG05GG52G at Caltech.
\end{ack}

\section*{References}
\bibliographystyle{iopart-num}
\bibliography{junk}

\end{document}